\documentstyle[aps,twocolumn,floats,prl]{revtex}
\begin{document}
\title{General analysis of self-dual solutions for the 
Einstein-Maxwell-Chern-Simons theory in $(1+2)$ dimensions}

\author{T. Dereli and Yu.N.\ Obukhov\footnote{On leave from: Department 
of Theoretical Physics, Moscow State University, 117234 Moscow, Russia}}
\address
{Department of Physics, 
Middle East Technical University, 06531 Ankara, Turkey}

\maketitle

\begin{abstract}
The solutions of the Einstein-Maxwell-Chern-Simons theory are studied
in $(1+2)$ dimensions with the self-duality condition imposed on the
Maxwell field. We give a closed form of the general solution which is
determined by a single function having the physical meaning of the
quasilocal angular momentum of the solution. This function completely
determines the geometry of spacetime, also providing the direct 
computation of the conserved total mass and angular momentum of the
configurations. 
\end{abstract}

\pacs{PACS no.: 04.50.+h; 04.20.Jb; 03.50.Kk}



The $(1+2)$-dimensional general relativity has attracted considerable 
attention recently (see, e.g., \cite{carlip} and references therein).
This is explained by two main reasons. Firstly, since the discovery of
the BTZ black hole solutions \cite{btz}, the three-dimensional gravity
became a helpful laboratory for the study of geometrical, statistical
and thermodynamics properties of black holes. Secondly, the quantization
of these models may give new insights into the general quantum gravity problem.

A number of generalizations of BTZ solution to the case of nontrivial 
electromagnetic field source were developed previously 
\cite{magsol,Chan,koi,fern}. The aim of our present paper is to give a 
new general analysis of the self-dual Einstein-Maxwell solutions in 
three dimensions. 

The Lagrangian 3-form,
\begin{equation}
L = {1\over 2}\,{\cal R}\,\ast\!1  - \lambda\ast\!1  - 
{1\over 2}F\wedge\ast F - {\mu\over 2}A\wedge F,\label{Ltot}
\end{equation}
contains the Einstein-Hilbert term, the cosmological constant $\lambda$, 
and the standard Maxwell field $F=dA$ Lagrangian along with the Chern-Simons
term with the coupling constant $\mu$, \cite{djt}. Variation of $L$ with 
respect to the coframe field $\vartheta^\alpha$ and the electromagnetic 
potential $A$ yields the system of field equations:
\begin{eqnarray}
G_{\alpha\beta}\,\ast\!\vartheta^\beta + \lambda\ast\!\vartheta_\alpha 
&=& \Sigma_\alpha,\\  d\ast\! F + \mu\,F &=& 0.
\end{eqnarray}
Here $\Sigma_\alpha = {1\over 2}\left[(e_\alpha\rfloor F)\wedge\ast F -
F\wedge e_\alpha\rfloor\ast\! F\right]$ is the Maxwell field 
energy-momentum 2-form, and $G_{\alpha\beta}$ is the Einstein tensor.

In the study of the ``spherically''-symmetric solutions, we choose the
local coordinates $(t,r,\phi)$ and make the general ansatz for the coframe 
1-form,
\begin{equation}
\vartheta^0 = f\,dt,\quad \vartheta^1 = g\,dr,\quad 
\vartheta^2 = h\,(d\phi + a\,dt),\label{coframe}
\end{equation}
and for the Maxwell field  
\begin{equation}
F = E\,\vartheta^0\wedge\vartheta^1 + B\,\vartheta^1\wedge\vartheta^2,
\end{equation}
Here $f, g, h, a$ and $E, B$ are the functions of the radial coordinate $r$.

Without any loss of generality it will be convenient to absorb the metric
function $g(r)$ by the simple redefinition of the radial coordinate:
\begin{equation}
\rho = \int g(r)\,dr \qquad ({\rm hence}\ \vartheta^1 = d\rho).
\end{equation}
>From now on, the derivatives w.r.t. new coordinate $\rho$ will be denoted by prime.

After all these preliminaries, the Einstein field equations read explicitly
\begin{eqnarray}
-\,{1\over 2}\,\beta' - \beta\gamma &=& - EB,\label{eq1}\\
\gamma' + \gamma^2 + {1\over 4}\,\beta^2 + \lambda &=& 
-\,{1\over 2}\,(E^2 + B^2),\label{eq2}\\
\alpha' + \alpha^2 - {3\over 4}\,\beta^2 + \lambda &=&
{1\over 2}\,(E^2 + B^2),\label{eq3}\\
-\,\alpha\gamma - {1\over 4}\,\beta^2 - \lambda &=&
{1\over 2}\,(E^2 - B^2),\label{eq4}
\end{eqnarray}
and this system is supplemented by the (modified) Maxwell equations:
\begin{eqnarray}
-\,B' - \alpha B + \beta E + \mu E &=& 0,\label{eq5}\\
-\,E' - \gamma E + \mu B &=& 0.\label{eq6}
\end{eqnarray}
Here we introduced the functions
\begin{equation}
\alpha = {f'\over f},\quad \beta = {a'h\over f},\quad
\gamma = {h'\over h}.\label{abc}
\end{equation}
which actually describe the Levi-Civita connection coefficients.
The remarkable feature is that the complete Einstein-Maxwell system 
(\ref{eq1})-(\ref{eq6}) involves no metric functions (i.e., $f,g,h,a$),
but only the connection combinations $\alpha, \beta, \gamma$. 

Let us assume ``self-duality'' of the electromagnetic field:
\begin{equation}
E = k\,B,\qquad {\rm with}\qquad k^2 = 1.\label{self}
\end{equation}
Substituting this into (\ref{eq5})-(\ref{eq6}) and (\ref{eq4}), 
we find that the two unknown functions are expressed in terms of the
third:
\begin{equation}
\alpha = {k\over 2}\,\beta + \ell ,\quad
\gamma = -\,{k\over 2}\,\beta + \ell. \label{aga}
\end{equation}
Here we denote $\ell :=\pm\sqrt{-\lambda}$.

Taking into account the algebraic relations (\ref{self}) and (\ref{aga}),
we are left with two essential equations for determining the functions 
$\beta$ and $B$. Explicitly, the equations (\ref{eq1}) and (\ref{eq6}) 
are reduced to
\begin{eqnarray}
\beta' - k\,\beta^2 + 2\ell \,\beta &=& 2k\,B^2,\label{s1}\\
(B^2)' - k\,\beta\,B^2 + 2\ell \,B^2 &=& 2k\mu\,B^2.\label{s2}
\end{eqnarray}
This system of nonlinear coupled equations is simplified with the
help of the substitution
\begin{equation}
\beta = {1\over\omega}, \quad 
B^2 =  {k\over 2\omega}\,{\varphi'\over \varphi},\label{uphi}
\end{equation}
which yields for the new functions $\varphi$ and $\omega$ the linear
equations:
\begin{eqnarray}
&& \varphi'' = 2k\mu\,\varphi'.\label{varphi1}\\
&& \omega' + \left({\varphi'\over \varphi} - 2\ell\right)
\,\omega + k = 0.\label{omega1}
\end{eqnarray}
Multiplying (\ref{omega1}) by $\varphi\,e^{-2\ell\rho}$,
we easily obtain the general solution 
\begin{equation}
\omega = {k\Omega\over \varphi\,e^{-2\ell\rho}},\quad {\rm with}\quad
\Omega := c_0 - \int\limits^\rho d\tilde{\rho}\,\varphi(\tilde{\rho})
\,e^{-2\ell\tilde{\rho}}.\label{om-gen}
\end{equation}
Note that in fact it is not necessary to know the explicit form of
$\varphi$ when solving (\ref{omega1}). At the same time, of course,
the equation (\ref{varphi1}) is straightforwardly integrated. Depending
on $\mu$, it admits two solutions: 
\begin{eqnarray}
\varphi &=& \rho + \rho_0,\quad {\rm when}\quad \mu =0,\label{var0}\\
\varphi &=& 1 + u_0e^{2k\mu\rho},\quad {\rm when}\quad \mu \neq 0.
\label{var1}
\end{eqnarray}
Here $c_0, \rho_0, u_0$ are integration constants. It is worthwhile
to note that an overall constant factor is irrelevant for $\varphi$
because this function appears everywhere only through the ratio (\ref{uphi}).

Quite remarkably, however, we will not need the explicit form of
$\varphi$ till the very end of our analysis. Such a formulation is
extremely convenient since it makes it possible to treat the cases of 
standard Maxwell theory with $\mu = 0$, and the Maxwell-Chern-Simons 
with $\mu\neq 0$ simultaneously. 

It remains to integrate the equations for the metric functions (\ref{abc}).
This is straightforward, and using (\ref{om-gen}) in (\ref{aga}), we find:
\begin{eqnarray}
f &=& f_0\,e^{\ell\rho}\,\Omega^{-\,{1\over 2}},\label{f}\\
h &=& h_0\,e^{\ell\rho}\,\Omega^{1\over 2},\label{h}\\
a &=& {k\,f_0\over h_0}\,\Omega^{-1} - a_0.\label{a}
\end{eqnarray}
For completeness, the magnetic field reads:
\begin{equation}
B^2 = {\varphi'\over 2}\,e^{-2\ell\rho}\,\Omega^{-1}.\label{B2}
\end{equation}
Here $f_0, h_0, a_0$ are integration constants.

The first main result which we learned in our study, is that a general 
``spherically''-symmetric (rotating, for nontrivial $a$) solution 
\begin{equation}
ds^2 = -\,\left(\vartheta^0\right)^2 + 
\left(\vartheta^1\right)^2 + \left(\vartheta^2\right)^2 \label{met}
\end{equation}
of the Einstein-Maxwell (with or without Chern-Simons term) field
equations is always represented solely in terms the function $\varphi$.

Because of such an important role played by $\varphi$, it would be
interesting to find out its physical meaning. The latter is revealed in
the analysis of the quasilocal mass and angular momentum which 
characterize our general solution. 

We refer the reader to \cite{brown} for a comprehensive discussion of 
the conserved quantities for gravitating systems within the framework
of Hamiltonian formulation of general relativity theory. As a first
step, let us use the coordinate freedom and replace $\rho$ by a new
radial coordinate defined by
\begin{equation}
r = h(\rho).\label{rh}
\end{equation}
Then a nontrivial metric function $g$ will reappear in the coframe
(\ref{coframe}) [and hence in the metric (\ref{met})]. Using (\ref{h})
we find explicitly
\begin{equation}
g = {d\rho \over dr} = \left(\ell\,r - 
{h_0^2\over 2r}\,\varphi\right)^{-1}.\label{g}
\end{equation}
Now we can write the quasilocal angular momentum at a distance $r$,
which reads
\begin{equation}
j(r) = {g^{-1}\,r^3\over f}\,{da\over dr},
\end{equation}
in our notations. Using (\ref{f})-(\ref{a}) and (\ref{rh})-(\ref{g}),
we find 
\begin{equation}
j(r) = k\,h_0^2\,\varphi.\label{angmom}
\end{equation}
Clearly, one should invert (\ref{rh}) and use $\rho =\rho(r)$ in 
(\ref{angmom}), or alternatively, one can consider the angular
momentum $j$ as a function of $\rho$.

The quasilocal energy is given, in our notations, by the difference
\begin{equation}
E(r) = g_0^{-1} - g^{-1},
\end{equation}
where the first term describes the contribution of the background 
``empty'' spacetime. The latter, as usually, is given by $g_0^{-1} =
\ell\,r$. Making use of (\ref{g}), we obtain explicitly
\begin{equation}
E(r) = {h_0^2\over 2r}\,\varphi = {k\over 2r}\,j(r).\label{energy}
\end{equation}
Finally, the quasilocal mass is determined by the expression
\begin{equation}
m(r) = 2\,f\,E(r) - j\,a.
\end{equation}
Substituting (\ref{f}), (\ref{a}), (\ref{angmom}) and (\ref{energy}),
we arrive at the result:
\begin{equation}
m(r) = a_0\,j(r).
\end{equation}

We thus have demonstrated that the function $\varphi$, which determines 
the spacetime geometry via (\ref{om-gen}) and (\ref{f})-(\ref{a}), is 
also determining {\it all} the quasilocal quantities of the gravitating 
system: its energy, mass and angular momentum. They turn out to be 
proportional to each other, describing a sort of extremal configuration. 
Because of the relation (\ref{angmom}), one can say that the angular momentum
$j(r)$ underlies the construction of self-dual Einstein-Maxwell equations: 
given this function, the metric and electromagnetic field are described
by (\ref{f})-(\ref{B2}) with $j(r)$ inserted. 

The total angular momentum and mass are defined by the limits 
$J := j\vert_{r\rightarrow\infty}$ and $M := m\vert_{r\rightarrow\infty}$, 
respectively. In order to find these quantities, one does not need to obtain 
the explicit exact form of the inverse coordinate transformation $\rho(r)$ 
from (\ref{rh}). It is sufficient to investigate the approximate behaviour 
of $\varphi(r)$ and $\Omega(r)$ for large values of $r$, which is always 
clear directly from the inspection of (\ref{varphi1})-(\ref{om-gen}).

In particular, one can immediately verify that the limiting value $\Omega
\vert_{r\rightarrow\infty}$ is equal either infinity or $c_0$, depending
on the values of $\mu$ and $\ell$. Consequently, the integration constant
$a_0$ should be equal either $0$, or ${kf_0\over h_0c_0}$, providing the
required asymptotic vanishing of the metric function $a(r)$. Correspondingly,
one finds that the quasilocal mass $m$ vanishes for many configurations. 

The quasilocal angular momentum $j(r)$ (or the function $\varphi(r)$) 
diverges, in general, for $r\rightarrow\infty$. However, the direct 
analysis of (\ref{varphi1})-(\ref{om-gen}) shows that $J$ is finite for
all the solutions with $k\mu<0$. Actually, there are two large classes
of such configurations: $(A)$ $k\mu<0, \ell =0$, then $J=kh_0^2$ and $M=0$, 
and $(B)$ $k\mu<0, \ell> 0$, then $J=kh_0^2$ and $M= {f_0h_0\over c_0}$. 
Imposing the standard asymptotic condition $f^2g^2\vert_{r\rightarrow\infty}
=1$, one finds $a_0 = k\ell$, and thus the solutions of the class $(B)$
are all characterized (irrespectively of the value of the Chern-Simons
coupling constant $\mu$) by $M^2 = \ell^2\,J^2$. This class also contains
the extremal BTZ solution, as a particular case when the electromagnetic
field is absent. [The general non-extremal BTZ solution cannot be recovered 
because of the algebraic relations (\ref{aga}) which necessarily hold 
for the self-dual electromagnetic field].

Summarizing, we have obtained a general solution of the 
Einstein-Maxwell-Chern-Simons theory in $(1+2)$ dimensions which covers
all the particular cases studied previously. The form of the solution 
(\ref{f})-(\ref{B2}), (\ref{om-gen}) is transparent and easy to analyse:
everything is determined by a single function $\varphi(\rho)$ which has
a clear physical meaning as the quasilocal angular momentum of the
gravitational field configuration. The computation of the total mass and 
angular momentum is straightforward and it involves only the analysis of 
the asymptotic behaviour of $\varphi$. 

The authors are grateful to TUBITAK for the support of this research. 
Y.N.O. is also grateful to the Department of Physics, Middle East 
Technical University, for the warm hospitality.


\end{document}